\documentstyle[12pt,psfig]{article}
\newcommand{\be}{\begin{equation}}
\newcommand{\ee}{\end{equation}}
\newcommand{\bea}{\begin{eqnarray}}
\newcommand{\eea}{\end{eqnarray}}
\newcommand{\ba}{\begin{array}}
\newcommand{\ea}{\end{array}}

\def\simlt{\stackrel{<}{{}_\sim}}
\begin{document}
\begin{titlepage}
\vspace*{-2.5cm}
\begin{flushright}
\vspace{-0.3cm}
{\normalsize CERN-TH/95-53}\\ 
{\normalsize OHSTPY-HEP-T-95-008}\\
\end{flushright}
\vspace{0.5cm}
\begin{center}
{\Large \bf Fermion Masses, Mixing Angles and \\ 
\vskip 0.1cm
Supersymmetric  SO(10) Unification  } \\
~\\
~\\
{\bf M. Carena,  S. Dimopoulos and C.E.M. Wagner} \\
 ~\\
 Theory Division, CERN, CH 1211, Geneva, Switzerland. \\
~\\
{\bf S. Raby}$^{\dagger}$\\
~\\
Department of Physics, The Ohio State University, \\
Columbus, OH 43210, U.S.A.
~\\
~\\
\end{center}
\begin{quote}
\begin{center}
{\bf Abstract}
\end{center}
We reanalyse the problem of fermion masses in supersymmetric 
SO(10) grand unified models. In the minimal model, both low 
energy Higgs doublets belong to the same  {\bf{10}} representation of 
SO(10) implying the unification  not only of  the gauge but also
of the third generation Yukawa couplings. These models  predict  
large values of $\tan\beta \sim 50$. In this paper we study the effects of
departing from the minimal conditions in order to see if we can find models 
with a reduced  value of $\tan\beta$.  
In order to maintain predictability, however,
we try to do this with the addition of only one new parameter.  We still assume
that the fermion masses arise from interactions of the spinor representations
with a single ${\bf 10}$ representation, 
but this ${\bf 10}$ now only contains a
part of the two light Higgs doublets.  This enables us to introduce one new
parameter  $\omega = \lambda_b/\lambda_t$.   For values 
of $\omega \ll 1$ we can
in principle reduce the value of $\tan\beta$. In fact, $\omega$ is an overall
factor which multiplies the down quark and charged 
lepton Yukawa matrices.  Thus
the theory is still highly constrained.  
We show that the first generation quark
masses and the CP-violation parameter 
$\epsilon_K$ are sufficient to yield strong
constraints on the phenomenologically allowed models. In the end, we find that
large values of $\tan\beta$ are still preferred.
\end{quote}
\vskip 0.5cm 
\begin{flushleft} 
{\normalsize CERN-TH/95-53}\\
{\normalsize OHSTPY-HEP-T-95-008}\\ 
{\normalsize March 1995}\\
~\\ 
$^{\dagger}$ Partially supported by DOE grant
DOE/ER/01545-647 \\

\end{flushleft} 
\end{titlepage}
\setcounter{footnote}{0}
\newpage 
\noindent
\section{Introduction} The Standard Model describes with a
great degree of precision the observed elementary particle interactions.  It 
provides, however, no answer to the fundamental questions about the origin of
the  gauge group $SU(3)_c\times SU(2)_L \times U(1)_Y$, 
the structure  of fermion
masses and mixing angles and  their quantum numbers.  Grand Unified Theories
(GUTs) have the power to fill the gap 
between theory and experiment \cite{GG}.   
Indeed,  within this framework the low energy group   proceeds   from the
spontaneous  breakdown of a single compact group.  The simplest   and most
attractive grand unified theories are based on the unitary group $SU(5)$ or the
orthogonal group  $SO(10)$.  Remarkably,  all 
low energy fermion quantum numbers
find a natural explanation within these theories.  For instance,   the fifteen
Weyl fermions in a Standard Model family, with their correct quantum
numbers under the standard model gauge group, are contained in a
${\bf 10}$ and a ${\bf 5}$  representation of $SU(5)$.  Most notably, they
are contained in a single spinor representation of $SO(10)$, the extra
state having the quantum numbers of a right handed neutrino 
and leading
therefore  to the possibility of including neutrino masses in a natural way.  

If the grand unified group breaks at very high energies to the standard model
gauge group, an essential requirement is that the theory should be 
supersymmetric \cite{DGR}.  
Not only does supersymmetry stabilize the hierarchy 
between the grand unified scale and the weak scale, but also the predictions
coming from gauge coupling unification within supersymmetric theories are
in remarkably good  agreement with the precise measurements of the weak 
mixing angle performed at LEP \cite{unif}-\cite{CPW}.  
Moreover, supersymmetry provides the
natural framework for the construction of a theory of quantum gravity,
and hence for the unification of all forces observed in nature. 
Supersymmetric grand unified theories provide also a simple theoretical
framework for the understanding of fermion masses.  The condition of
bottom--tau Yukawa coupling unification implies, for instance, a large
value of the top quark Yukawa coupling at the grand 
unification scale \cite{Ramond},\cite{DHR},
the low energy value of the top quark mass being
governed, in general,  by the
infrared fixed point structure of the theory \cite{IR}-\cite{Dyn}.  
Hence, supersymmetric
GUTs provide an understanding of the large value of the top quark
mass \cite{LP},\cite{CPW},\cite{BABE},\cite{BCPW}.   
Moreover, in the minimal SO(10) model, the three Yukawa couplings
of the third generation unify at the grand unification scale. This
yields predictions  not only for  the top quark mass, but also for the
ratio of  Higgs vacuum expectation values, $\tan\beta$, which becomes
naturally large \cite{somp}.   
Large values of $\tan\beta$ are also associated with
large corrections to the bottom mass \cite{Hall},\cite{Ralf}, 
which  depend on the
supersymmetric spectrum and which should be computed in a consistent way
in order to obtain phenomenologically correct predictions for the top quark
mass \cite{wefour}.

\section{Minimal SO(10) models}
The hierarchy between the third and the first and second generation quark
masses, as well as the 
inter--generation mixing angles, may be explained
by assuming that only the third generation quarks couple to the ${\bf 10}$
of Higgs by renormalizable interactions, while the other  mass terms are
induced through
higher order operators. A systematic search for this class of
models within the framework of 
mimimal $SO(10)$  was done in Ref. \cite{ADHRS},
under the assumption that the model includes only three spinor
representations, containing the three low energy families,  
a few extra
heavy spinor representations, the ${\bf 10}$ Higgs multiplet and some
 ${\bf 45}'s$ ,
necessary for the correct breakdown of the gauge symmetry and for
the generation of the fermion mass operators.
All higher order operators are of the form

\begin{equation}
O_{ij} = 16_i  \;\;  {M_G^k ~45_{k+1} \cdots 45_m \over M_P^l
{}~45_X^{m-l}} \;\;
  10  \;\; {M_G^n ~45_{n+1} \cdots 45_p \over M_P^q
{}~45_X^{p-q}} \;\;
 16_j ,
\label{eq:oper}
\end{equation}
where the $45$ vevs in the numerator can be in any of the 4 directions,
${\bf X, Y, B-L, T_{3R}}$ (discussed below) and the $45$ in the denominator can
only be in the ${\bf X}$ direction which breaks $SO(10)$  down to the subgroup
$SU(5)\times U(1)_X $.  This occurs at a scale $M_{10}$ which is assumed to lie
between the GUT scale $M_G \sim 10^{16} GeV$ and the Planck scale $M_P$. 

 The adjoint {\bf 45}'s may be labelled according to  the direction of their
vacuum expectation values. There are four special directions \cite{ADHRS}. 
The $X$  direction,
necessary for the breakdown of $SO(10)$ to $SU(5) \times U(1)_X$
at the scale $M_{10}$. The ${\bf 45_X}$ in the 
denominator can arise when integrating out 
heavy ${\bf 16}$ and ${\bf \overline{16}}$ states 
with mass from the ${\bf 45_X}$ vev.  Of course, 
this only makes sense if $M_{10} > M_G$.  Other  directions are the
$Y$  and $B - L$, 
which break $SU(5)$ to the Standard Model gauge group.
The presence of the latter is required for a natural solution of the 
doublet--triplet splitting problem in this theory \cite{DW}.  
Finally, there is another,
linearly dependent direction, ${\bf T_{3R}}$,  
which, as we shall explain below,
may be useful to achieve low values of $\tan\beta$ within this model.

 Taking into account the experimental constraints on the 
lowest generation fermion masses and the 
Cabibbo-Kobayashi-Maskawa (CKM) mixing
angles,  the authors of Ref. \cite{ADHRS} identified nine potentially 
acceptable models, in which the up and down quark  and lepton 
mass matrices  are of the form,
\begin{eqnarray}
\lambda_a = \left( 
\begin{array}{c c c}
0 & z^{'}_a C & 0 \\
z_a C & y_a E e^{i\phi} & x^{'}_a B \\
0  & x_a B & A 
\end{array}
\right),
\end{eqnarray}
where  $z_a$, $z^{'}_a$, $x_a$, $x^{'}_a$ and $y_a$ are Clebsch factors, while
$A$, $B$, $C$, $E$ and $\phi$  are 
arbitrary parameters, which respect the hierarchy $A \gg B,E \gg C$ and
must  be adjusted in order to
obtain predictions in agreement with the present data. 
The Higgs sector provides an additional free paramater, which is the
ratio of vacuum expectation values, $\tan\beta$. Using 
the presently best known 
low energy parameters $m_e$, $m_{\mu}$, $m_{\tau}$, $m_c$,
$m_b$ and $|V_{cd}|$ as input,  the values of  $M_t$, $\tan\beta$,
$|V_{cb}|$,  $|V_{ub}|$, $ m_u$,  $m_d$, $m_s$ 
and the CP-odd Jarlskog invariant 
$J$ \cite{Jarlskog}  are predicted (we shall denote physical and
running masses by capital and small letters, respectively).   
This leads, hence, to eight low energy predictions, which 
should be compared with the present experimental values.

There are several properties, which are shared by all these models. 
First of all, they maintain the Georgi--Jarlskog 
relation \cite{GJ} of the
$y_a$ Clebsch factors: $|y_e| : |y_d| : |y_u|\equiv 3 : 1 : 0$.
This relation of Clebsch factors appears in a natural way, for example,
through the operator
\begin{equation}
O_{22} = 16_2 \ {45_X \over M} \ 10 \ {45_{B-L} \over 45_X} \ 16_2,
\end{equation}
and it is important in order to derive correct predictions
for the first two generations of  quark masses. In fact, 
\begin{equation}
\frac{m_s}{m_d} \simeq \left( \frac{y_d}{y_e} \right)^2 
\frac{m_{\mu}}{m_e} \left| \frac{z_e z^{'}_e}{z_d z^{'}_d}
\right|.
\label{eq:ratio}
\end{equation}
Hence, as long as the equality $z_e z^{'}_e = z_d z^{'}_d$ holds,  the ratio
of lepton and quark masses is in good quantitative agreement with
the observed experimental values.  

Another important property of these models is
the unification of the three Yukawa
 couplings of the third generation and, in particular, the unification
of the bottom and top Yukawa couplings,
which requires large values of $\tan\beta$.  
 Such large values of $\tan\beta$
are associated with three effects:
\begin{enumerate}
\item potentially large corrections to the down quark mass
matrix  (these radiative corrections are discussed in detail 
in the appendix);
\item  with some fine-tuning of GUT scale soft SUSY breaking 
parameters in order to obtain 
radiative electroweak symmetry breaking at the weak scale.  
The range of parameters which satisfy the second 
constraint ({\em when universal scalar masses are imposed at M$_G$}), 
in fact, requires the corrections to down quark masses to be large, and
\item the proton decay rate resulting from dimension 
5 baryon violating  interactions is enhanced. 
\end{enumerate}
  It has recently been shown that the first two consequences of 
large $\tan\beta$ are  ameliorated when the constraint of 
universal scalar masses is removed\cite{nonunl}.  
The corrections to the down quark masses can be small and 
the amount of fine-tuning is greatly reduced.  The problem of an enhanced 
proton decay rate is unaffected.
On the other hand, these strong constraints become weaker for smaller
values of $\tan\beta$.   It becomes an
important question whether the prediction for large $\tan\beta$ can be altered
without destroying the predictability of the theory.  

\section{Trying to reduce $\tan\beta$ in minimal SO(10) models}
 
Lower values of $\tan\beta$ can easily be achieved by assuming that
only one ${\bf 10}$ of Higgs couples to fermions, but this ${\bf 10}$
contains only a piece of the two Higgs doublets, the other
components coming, for instance, from an additional ${\bf 10}$.
The overall effect is
to multiply the down and lepton mass matrices by a factor
$\omega$, which is the ratio of the relative components of the
two Higgs doublets in the ${\bf 10}$ which couples to fermions. The 
minimal model would hence be obtained for $\omega = 1$.

Such a situation can come about as follows: Consider the superpotential
\begin{equation}
W = {\bf 10 \; 45_{B-L} 10^{'}}  + 
\left[ M_1 {\bf 10^{'}} + \left( M_2 + {\bf 45_X} \right) 
{\bf 10} \right] {\bf 10}^{''}
\end{equation}
where $M_1$ and $M_2$ are of order $M_{GUT}$; ${\bf 10}$,
${\bf 10^{'}}$ and
${\bf 10^{''}}$ are decouplets and only ${\bf 10}$ participates in the
fermion mass operators.

The first term in $W$ implements the Dimopoulos-Wilczek 
mechanism \cite{DW} and yields 4 light doublets:
{\bf 2}, ${\bf \bar{2}}$, ${\bf 2^{'}}$ and ${\bf \bar{2}^{'}}$;
the color triplets get a mass of order $M_{GUT}$.
The second term gives a mass to a linear combination of
{\bf 2} and ${\bf 2^{'}}$ (by pairing it with ${\bf \bar{2}^{''}}$)
and a different linear combination of ${\bf \bar{2}}$ and
${\bf \bar{2}^{'}}$ (by pairing it with $2^{''}$).
Explicitly, the light states are given by
\begin{equation}
2_L = \frac{ M_1 \; 2 - \left( M_2 + {\bf v} \right) 2^{'}}{\sqrt{M_1^2
+ \left(M_2 + {\bf v} \right)^2}}
\end{equation}
and
\begin{equation}
\bar{2}_L = \frac{ M_1 \; \bar{2}  - 
\left( M_2 - {\bf v} \right) \bar{2}^{'}}
{\sqrt{M_1^2 + \left(M_2 - {\bf v} \right)^2}} ,
\end{equation}
where ${\bf <45_X> = v} \times X$, with $X = +$ $(-)$ when it acts
on the ${\bf 5}$  (${\bf \bar{5}}$)
 of a ${\bf 10}$ representation,
respectively.  Since ${\bf 2}$ couples to the up quarks and
${\bf \bar{2}}$ couples to the down quarks, in this example we have
\begin{eqnarray}
h_t & = & \lambda \frac{M_1}{\sqrt{M_1^2 + 
\left(M_2 + {\bf v} \right)^2}} ,
\nonumber\\
h_b & = & \lambda \frac{M_1}{\sqrt{M_1^2 + 
\left(M_2 - {\bf v} \right)^2}} ,
\end{eqnarray}
and
\begin{equation}
\omega = \frac{\sqrt{M_1^2 + \left(M_2 + {\bf v} \right)^2}} 
{\sqrt{M_1^2 + \left(M_2 - {\bf v} \right)^2}}.
\end{equation}
Notice that, in this simple example, 
$M_1$ (or $M_2 \pm {\bf v}$) cannot be too small, 
or else  a pair of light triplets {\bf 3} and ${\bf \bar{3}}$ 
would appear in the spectrum, affecting the prediction for
$\sin^2 \theta_W$. Hence, $\omega$ cannot become
too small in this case.

From now on, we shall discuss the consequences of
the departure from the minimal conditions, taking values of
$\omega$ lower than one. Values of $\omega$ lower than one
decreases the bottom to top Yukawa coupling ratio but still requires 
bottom--tau Yukawa coupling unification.

\subsection{SO(10) models with moderate values of $\tan\beta$  --- 
The second and third generations}
 
We have introduced the parameter $\omega \le 1$ in an
attempt to lower $\tan\beta$. In this section  
we discuss the results for the second and third generations with the
additional parameter $\omega$.  In general, taking into account variations of
$\omega$ and reasonable assumptions on radiative corrections to down quark
masses, we find that, in order to avoid a very
heavy top quark, with mass larger
then 190 GeV, the value of $\tan\beta$ should be either larger than 20
or very close to 1.  

As a general feature,
in order to obtain unification of the bottom and 
tau Yukawa couplings, the third
generation Yukawa couplings must partially compensate the strong gauge coupling
renormalization group effects.  For $\omega = 1$, this is partially achieved by
large values of the bottom  Yukawa coupling. Indeed, the relation between the
bottom  quark and tau masses is given by 
\begin{equation} 
\frac{m_b}{m_{\tau}} =
{\bf G} \exp(-I_t - 3 I_b + 3 I_{\tau}), 
\end{equation} 
where {\bf G} includes
the $\omega$ independent, gauge coupling dependent factors, $I_a = \int
(h_a/4\pi)^2 dt$ with  $h_a$  the corresponding Yukawa coupling and $t =
\ln(Q/M_Z)$. In the following, we shall always assume that the right handed
neutrinos acquire large Majorana masses of order $M_{GUT}$ and hence decouple
from the renormalization group equations.  Although there is a partial
cancellation of the bottom and tau Yukawa 
coupling contributions at scales close
to the unification scale,  due to the factor 3 and the  relation $I_b >
I_{\tau}$ , the bottom contribution becomes important for $\omega = 1$. For 
values of $\omega < 1$, for which only the bottom and $\tau$ Yukawa couplings
unify,  the top Yukawa coupling must increase in order to 
compensate for the smaller contribution of the bottom 
Yukawa coupling.  For smaller values:  $\omega < 0.5$, associated
with moderate or small values of $\tan\beta$,  
and in the absence of 
supersymmetric threshold corrections, the top quark Yukawa coupling
must acquire large values at the grand unification scale, 
being driven towards its infrared fixed point value at low energies.
The convergence of the top quark mass to its fixed point value 
is naturally weaker for  $\omega \simeq 1$.

For  $\tan\beta \geq  5$, the fixed point value of the 
pole top quark mass reads
$M_t \simeq$ 190--210 GeV, which is somewhat large in comparison to the
current experimentally preferred value  
$M_t \simeq  180 \pm 12$ GeV \cite{CDFD0}. 
The convergence to the fixed point for moderate values of 
$\omega$ may  be softened by the presence of large
bottom mass corrections, which become
particularly relevant for values of $\tan\beta > 10 $.
For  values of $\tan\beta \leq 5$, the 
bottom mass corrections are generically small, but
the infrared fixed point value
of the top quark mass, $M_t \simeq  \sin \beta \times 200$ GeV, 
is lowered by the $\sin\beta$ factor (see Fig. 1).  
Indeed, values of $\tan\beta \simlt 3$ are
required, for the fixed point solution to be in the range of 
phenomenologically preferred values. As we shall discuss below,
these small values of $\tan\beta$  demand very small values of $\omega$.  

Figure 1 shows the dependence of the 
pole top quark mass on $\tan\beta$ 
(and also on $\omega$)  for 
three  different values of $\alpha_3(M_Z)$ and  
different values  of  the coefficient $K_c$ parametrizing
the bottom mass corrections, $\delta m_b = - m_b K_c  \tan\beta$.  
Values of $K_c \geq  0.005$ lead to significant
corrections to the predicted top quark mass values and, as was shown
in Ref \cite{wefour},  may appear in the presence of universal
soft supersymmetry breaking mass parameters at the grand unification
scale. We concentrate on positive values of $K_c$, since for
negative and large $K_c$ 
either the top quark mass is above its experimentally
preferred values or a Landau pole in the top quark Yukawa coupling
appears at scales below $M_{GUT}$. 
In Figure 1 we have chosen a representative value of
$m_b(m_b) = 4.15$ GeV. Larger (lower) values of $m_b$ 
within the experimentally allowed range $m_b = 4.25 \pm
0.25$ GeV, would 
lead to somewhat lower (larger) values of $M_t$ \cite{CPW}, 
without  changing the general properties of the solutions.

It is interesting to note that, for large values of $K_c$
and low values of $\alpha_3(M_Z)$, the top quark
mass predictions in model 6 differ from the ones 
obtained in model 9 for
the same values of $\omega$. This reflects the effect of the mixing
between the second and third generations on the predictions for 
the third generation masses. It is easy to prove that, although
this effect is generically small, the $\tau$ mass in model 9 receives
a significant correction due to the mixing, which for values of
$\alpha_3(M_Z) = 0.115$ and $K_c = 0.006$ becomes of order 
$15 \%$. Due to the condition of bottom-tau Yukawa unification,
large tau mass corrections also imply large variations in the
top quark mass predictions.

To summarize, we observe that
depending on the size of the one loop supersymmetric corrections
to the down quark masses,  successful top quark mass predictions may
be obtained for the minimal models with  $\omega$ = 1, but also for
 moderate and
small values of $\tan\beta$ (associated with moderate or very small
values of $\omega$).  It is hence important to know if the same is
true  for the first two generations of quark masses and mixing angles. 
Since the relation between the down quark and lepton masses
is only weakly dependent on $\omega$, we should mantain the
Georgi-Jarlskog relation even for values of $\omega$ different 
from one.  Moreover,
the values of $V_{cb}$ can also be successfully 
accomodated for lower values of $\omega$.  This can be easily seen writing
its dependence   in terms of the top and charm
quark masses,  
\begin{equation}
|V_{cb}| \simeq \chi \sqrt{ \frac{m_c}{ \eta_c m_t} } 
\exp \left(\frac{I_b - I_t}{2} \right)
\label{eq:Vcb}
\end{equation}
where $\chi = | x_d - x_u|/ \sqrt{|x_u x^{'}_u|}$,
and $\eta_a$ are the $\omega$ independent, renormalization 
group factors relating the running masses at the scale $M_Z$ with
the on--shell ones (for the $u$, $d$ and $s$ quarks, the scale of
definition of the running masses  is taken to be 1 GeV). For 
$\alpha_3(M_Z) \simeq 0.12$, 
for which the value of $\eta_c \simeq 2.2$, 
it follows that the phenomenolgically preferred values of
$|V_{cb}| = 0.040 \pm 0.005$, require values of 
$\chi < 1$ \cite{Partd}.
The different values of $\chi$ are the basis for the classification
of models performed in Ref. \cite{ADHRS}, where the best fit to the
data was achieved by two models:
model 6, with $x_u = x'_u =$ -4, $x_d = x'_d =$ -2/3 and $x_e = x'_e =$  
6, and model 9 with $x_u = x'_u =$ 1, $x_d =$ 1/9,
$x_e =$ 9 and $x'_d = x'_e =$ 1.  These models have
$\chi = 5/6$ and $8/9$,
respectively and both lead to somewhat large values of $V_{cb}$.
Model 4, with
$\chi = 2/3$,  
leads to a better prediction for $V_{cb}$, but yields
insufficient CP-violation.

For  lower values of $\omega$, the dependence of
Eq. (\ref{eq:Vcb}) on $I_b$ and $I_t$ is such, that the values of
$V_{cb}$ tend to decrease. Moreover, in the absence of down 
quark mass matrix corrections, for $\tan\beta \geq 4$,
the value of $V_{cb}$  decreases due to larger top quark mass
values, which, as we discussed before, may become too large
in comparison with the experimentaly preferred ones. As shown
in Fig. 1, lower values of the top quark mass may be obtained
through large bottom mass corrections. Lowering the top quark
mass enhances the value of $V_{cb}$, but 
the total effect of the down quark mass corrections on $V_{cb}$
cannot be  determined a priori; it 
depends on the relative size of the gluino corrections,
which affect the value of $V_{cb}$ due to their effect on
the predicted top quark mass value, and the chargino corrections, which  
modify not only the  top quark mass value through the bottom mass
corrections, but they have also a direct effect on
the CKM matrix elements \cite{SS} (see Appendix).

Fig. 2 shows the predictions for $V_{cb}$ 
for models 6 and 9, as a function of 
$\tan\beta$, for three different bottom mass corrections
and three different values of $\alpha_3(M_Z)$, under the
assumption that $(\delta m_b/m_b)^{\tilde{g}} = -3
(\delta m_b/m_b)^{ch}$ (which is reasonable in view of the
running of the soft breaking parameters and the structure of
the bottom mass corrections  when the squark mass matrices
are approximately  three by three block diagonal \cite{wefour}). 
We see that, independently of the bottom mass corrections,
the predictions for $V_{cb}$ may be significantly improved for
moderate values of $\omega$. Indeed, apart from the solutions
with $\tan\beta$ very close to one,  
$\omega =1$ leads to the 
largest values of  $V_{cb}$ for each fixed $\Delta m_b$ correction.
Observe as well that for  values of $\tan\beta \leq 2$, $V_{cb}$ 
increases, due to the lower values of $m_t$ appearing in this regime. 
Furthermore, for the present case,
for any fixed value of $\omega$, there is an effective cancellation
of the chargino and gluino-induced  one loop 
corrections to $V_{cb}$ and the total effect of the down quark
mass corrections on $V_{cb}$ is small. Consequently, since for a fixed 
value of $\omega$
large down quark mass corrections lower the value of $\tan\beta$, 
as can be seen from Fig. 2,
they also yield larger values of $V_{cb}$ for a given fixed value
of $\tan\beta$.  From Fig. 2 we also observe that 
the predictions for $V_{cb}$
improve for larger values of $\alpha_3(M_Z)$. In
fact, for moderate values of $\omega$, if large values of 
$\alpha_3(M_Z)$ and large bottom mass corrections are 
present, the predictions for $V_{cb}$ in model 6 may actually 
be below the preferred experimental values, but these solutions
are associated with values of $M_t$ which are generally too large.

From the  discussion above, we see that the second and third generation
fermion masses and mixing angles can be consistently described within 
an SO(10) GUT  with $\omega \leq 1$. However, as we shall
show in the following, the constraint coming from the predictions
for the first generation quark masses rule out values of $\omega \le
0.5$ within the minimal model.  In section 4 we
show how to overcome this difficulty at the expense of adding one new operator
and 2 more parameters, in addition to $\omega$.

\subsection{ The first generation}

The operator $O_{12}$ is necessary to achieve
acceptable predictions for the lowest generation quark masses.  Indeed,
within the minimal model, there is a 
\lq\lq unique" operator, 
\begin{equation}
O_{12} = 16_1  \left( \frac{45_X}{M_P} \right)^{n} 10 
\left( \frac{45_X}{M} \right)^{m} 16_2,
\label{eq:O12}
\end{equation} 
with $n = m = 3$, which yields  acceptable
ratios for the masses of the up, down and strange quarks.
This  operator
determines the equality of the Clebsch factors $z_a$ and $z^{'}_a$ 
and the ratios of the Clebsch factors $z_d/z_u = 27$ 
and $z_d / z_e = 1$
(the ratio of Clebsch factors $z_d/z_u$  
increases  by a factor 3 for each power 
of $45_X$).  In addition the ratio, appearing in Eq. (\ref{eq:ratio}), 
$(z_e z'_e)/(z_d z'_d) = 1$. The above
operator is of dimension ten, meaning that the absence of any lower 
dimensional operators should be insured by some symmetry of the theory.

For $\omega = 1$, one might think that the
large ratio of Clebsch factors, $z_d/z_u$,  arising from the above
relation, Eq. (\ref{eq:O12}),  is necessary in order to compensate the 
$\tan\beta$ $( \simeq m_t/m_b) $ 
dependence of the up--type quark masses with respect
 to the down--type quark ones.  It is
interesting to investigate then if lower values of
$\omega$, and hence of $\tan\beta$,  can  serve to relax the restrictions on 
the Clebsch factors and hence, to lower the dimensionality
of the above $O_{12}$ operator.   This, however, is not the case, 
as can be easily shown  considering  the relation
\begin{equation}
\frac{m_u}{m_d} = \frac{m_s}{m_c} 
\left(\frac{m_t}{m_b}\right)^2 \frac{ \eta_u \eta_c \eta_b^2}
{\eta_d \eta_s} \left| \frac{z_u z^{'}_u}{z_d z^{'}_d }\right|
\exp 4 (I_t - I_b).
\label{eq:mumd}
\end{equation}
From Eq. (\ref{eq:mumd}) it follows  that independent of
the source of the hierarchy between the top and bottom quark masses,
large ratios of Clebsch factors are necessary in order to obtain the
phenomenologically preferred values for the ratio of the up to
down quark masses, $0.2 \leq m_u/m_d \leq 0.8$ \cite{KML}. 
 The additional  dependence on the integral factors $I_b$ and $I_t$ 
does not help to lower this ratio. On the contrary,
since for lower values of $\omega$,
the integral factor $I_b$ decreases, while $I_t$ changes only slightly,
for the same values of the second and third generation quark masses
the ratio of the up to the down quark masses increases.
This means that, in order to keep phenomenologically allowed
values of the first generation quark masses, the ratio of the
Clebsch factors $z_d/z_u$ should actually increase, 
implying  that the dimensionality of the operator $O_{12}$ should 
correspondingly increase for lower values of $\omega$. 

It is therefore apparent that, keeping the  same operator structure
as before,  the range of possible values of $\omega$, that is to say
of $\tan \beta$, will be restricted.
Indeed, Figure 3 shows the dependence of the ratio $m_u/m_d$
as a function of $\tan\beta$ ($\omega$) for   models 6 and 9
and for different values of the down quark mass corrections, under
the same assumptions discussed for Fig. 1.
It follows that the down quark mass corrections help only marginally
in getting  phenomenologically allowed values for $m_u/m_d$, 
and  values of $\omega \simlt 0.5$ are disfavoured for
all these models. Indeed, for larger values of $\alpha_3(M_Z) 
\geq 0.12$, even larger values of $\omega$ are necessary in order to
achieve good predictions for the first generation masses.

Increasing by one the dimensionality of the operator
$O_{12}$  keeps
the equality  $z_e z^{'}_e = z_d z^{'}_d$, necessary to achieve 
the proper ratio of first and second generation quark and lepton 
masses, Eq. (\ref{eq:ratio}), but leads to  
wrong predictions for the Cabibbo angle \cite{ADHRS}.
Indeed, ignoring small factors, the Cabibbo angle is 
approximately given by
\begin{equation}
s_c \simeq \sqrt{ \frac{m_d}{m_s}} \sqrt{\left| 
\frac{z_d}{z^{'}_d} \right|}.
\label{eq:sc}
\end{equation}
Relaxing the equality $z_d = z^{'}_d$ by increasing by one 
the power of one of the $45_X$ in $O_{12}$
would  change $s_c$ by a factor $\sqrt{3}$,
what would lead to predictions in conflict 
with present data.

Therefore,  for low values of $\omega$,  if
 the dimension of the operator $O_{12}$, Eq. (\ref{eq:O12}),
is changed,  to obtain correct values for the ratio of the 
first and second generation quark masses, it
 should be increased by two units.
Once more, however,  the variation in the dimensionality of
this operator has an additional effect, which is
related to the behaviour of the Jarlskog CP-odd invariant
$J  = Im[ V_{ud} V_{tb} V_{td}^{*} V_{ub}^*]$. 
Ignoring again small, inessential factors, it is straightforward to show 
that
\begin{equation}
J \simeq \chi^2 \frac{| z_u z_d|}{|z_e z_e^{'}|} 
\left| \frac{y_e}{y_d} \right|  \frac{m_e}{m_{\tau}}
\exp\left(2 I_t + 2 I_b - 3 I_{\tau}\right).
\end{equation}
Thus, increasing the dimension of the operator $O_{12}$ in two units
implies a decrease in the CP--odd invariant $J$ in a factor three.   
Since the observed CP--violation in the $K$ system is well described
by models 6 and 9 before the modification of the operator $O_{12}$,
a factor three suppression of the Jarlskog invariant   
would imply that the amount of CP--violation associated with the 
Cabibbo-Kobayashi-Maskawa is insufficient to explain the experimental
data.  The possible variations of $V_{cb}$ (or equivalently 
of $V_{td}$) due to supersymmetric threshold corrections
in the down quark sector, which we have discussed above, 
are not sufficiently large to  compensate this type of effect. 
Numerically, we observe the effect of increasing the dimension of the
operator $O_{12}$ 
through the prediction for the bag parameter $B_K$ \cite{epsK},
\begin{equation}
B_K \simeq  \epsilon_K \frac{|z_e z^{'}_e|}{\sqrt{|z_d z^{'}_d|} |z_u|}
\frac{m_{\tau}}{m_e} \sqrt{\frac{m_d}{m_s}}  \frac{ \exp(2 I_t + 2I_b -
3 I_{\tau})}{\chi^2 \; \sin\phi}.
\end{equation}
which tends to  be larger than one in all models, 
and,  hence, unacceptable since the phenomenologically
preferred values  are $B_K = 0.8 \pm 0.2$
\cite{BK},\cite{Partd}.

The general conclusion of this study is that, keeping the same
operator structure as in Ref. \cite{ADHRS}, values of $\omega \leq 0.5$
cannot be accomodated, without spoiling the predictions for either the
first generation quark masses,  the Cabibbo angle 
or the CP-violation sector of the theory.
Hence, the preferred value of $\omega \simeq 1$ restricts us to
be close to the minimal $SO(10)$ model  
and the values of
$\tan\beta$ and $M_t$ which lead to acceptable predictions are also 
quite restricted (see the discussion in section 3.1).  
In the next section we show
that it is still possible to obtain acceptable predictions for
the first generation with small values of $\tan\beta$.  However, this
solution requires the addition of one new operator and thus one more complex
parameter in addition to the free parameter $\omega$ discussed above.

\section{Extending minimal SO(10) and  $\tan\beta \sim 1$}

One could think of improving the agreement between the theoretical
predictions and the experimentally observed values  of the first
generation masses, or  the  $\epsilon_K$ parameter, 
by assuming very large  supersymmetric threshold corrections to these
variables. In Fig. 3 we have shown that if the down quark mass corrections
have the structure which naturally appears when the squark matrices are
block diagonal (see Appendix), only slight changes of the 
predictions for the first generation masses 
are obtained through such threshold corrections.
In the general case,  however, the squark mass matrices may be far from being
three by three block diagonal and  first generation 
down quark mass corrections,
proportional to the second or even third generation masses, as shown in the
appendix (Eq. (\ref{eq:bmcorr})), may be present. 

Persuing this direction
however opens up a pandoras box of new possibilities and new problems.
It is interesting to note that, if the
supersymmetry breaking is transferred to the observable sector through
gravitational  effects, a nontrivial  inter-generation squark mixing, generated
through renormalization effects at scales of the order of the grand unification
scale, is  unavoidable \cite{RHK},\cite{Geo},\cite{BarH}, \cite{DimH}.   
A reliable 
computation of this effect demands, however, the knowledge of the precise
physics beyond the grand unification scale. In general,  a large squark 
mixing would also involve large flavor changing neutral current effects. 
Barring unnatural cancellations, large flavor 
violations in the fermion sector can only be
consistent with the  experimental constraints on  flavor changing neutral 
currents and the neutron electric dipole moment if the characteristic scale
of the squark masses is larger or of order 1 TeV.  A large
squark mixing also implies significant couplings of these heavy squarks to
the Higgs sector of the theory (unless the third generation squarks do not
mix with the first and second generation ones), 
this will in turn imply a significant
fine tuning in order to preserve the stability  of the weak scale.
The presence of
large supersymmetric corrections to the $\epsilon_K$ 
parameter have similar
consequences. In this work, we assume the 
presence of a super GIM mechanism and avoid the discussion of non-universal 
squark and slepton masses at the GUT scale. Note that in order to reduce the
fine-tuning and large corrections associated 
with large values of $\tan\beta$ it
is only necessary to have non universal Higgs masses.

To improve the agreement between the theoretical
and experimental predictions for small values of $\tan\beta$,
a possible alternative is the modification  of the structure of the operators 
discussed above. Since, as shown section 3, low values of
$\omega$ are perfectly consistent with 
the second and third generation quark and
lepton masses and mixing angles, any modification should concentrate on
the form of the \lq\lq 12'' elements.  
In Ref. \cite{ADHRS}, it was argued
that if $O_{12}$ proceeds from  a single operator, its form is uniquely 
determined. This conclusion is based on the analysis of the associated
Clebschs and the relations given in  
Eqs. (\ref{eq:ratio}), (\ref{eq:mumd}) and (\ref{eq:sc}). However,  
since $O_{12}$ has a large dimension,
the relaxation of the assumption that the \lq\lq 12 " elements
come  from a single operator seems natural. If the effect of 
two operators
had to add in an unnatural way in order to lead to the correct
phenomenological predictions, the predictive power
of the theory would be spoiled. Therefore,
the additional operator should  not
modify the equality of $z_e z^{'}_e$ and $z_d z^{'}_d$ and should
give no relevant corrections to the ratio of $z_d/z^{'}_d$.   On the 
other hand, we want to modify the ratio of $m_u/m_d$ without
affecting the CP-odd sector in a relevant way.  
It is crucial to notice that
there is a very important difference between the dependence of
$J$ and that of $m_u/m_d$ on the Clebsch factors. While 
$m_u/m_d$ depends on the product $z_u z^{'}_u$,  
the CP-odd invariant $J$ depends on $z_u$, but is 
independent of $z_u^{'}$.  Hence, we are searching for an
operator which modifies $z_u^{'}$, leaving $z_u$ invariant.
There is only one combination of operators which fulfills 
all the above criteria, namely,
\begin{equation}
O_{12} = 16_1 (45_X)^n 10 (45_X)^n 16_2 +
K 16_1 (45_X)^m (45_{T_{3R}})^l 10 (45_X)^m  16_2
\end{equation}
where K is of order one. The predictions for  
$z_a$ and $z_a^{'}$ within this
framework are:
\begin{eqnarray}
z_u &=& 1, \;\;\;\; z^{'}_u = 1 + f
\nonumber\\
z_d &=& (-3)^n, \;\;\;\; z^{'}_d = (-3)^n + (-1)^l (-3)^m f 
\nonumber\\   
z_e &=& (-3)^n, \;\;\;\; z^{'}_e = (-3)^n + (-1)^l (-3)^m f,
\end{eqnarray}
where $f$ is the coefficient characterizing the relative
weights of the two contributions, and it is computable 
from $K$ and the vacuum expectation values above. 
For simplicity, we shall assume that $f$ is a real number.
In that case, for values of $f$ of order one and values of
$m$ smaller than $n$ by
at least two units, it is easy to see that the only prediction
which will  be modified considerably is $m_u/m_d$. One can 
therefore achieve 
low values of $\tan\beta$ with a correct prediction for $m_u/m_d$.
This demands very low values of $\omega$ and values of $f$ close
to $-1$. For instance, for model 6,
$m_b(m_b) \simeq$ 4.2 GeV,
and $\alpha_3(M_Z) \simeq 0.12$, the  value of $\omega$
which leads to $\tan\beta \simeq 1.5$ 
is as small as 0.004.  In this case,
values of $f \simeq - 0.8$, $n =3$  and $m = 0$  lead 
to good predictions for the CKM matrix and the quark masses.

\section{Conclusions} We have analysed the fermion mass problem within
the context of supersymmetric $SO(10)$ unification, studying
not only the minimal case, but also the departure from the
minimal conditions assuming that the fermion masses arise from
interactions of the spinor representations with a single 
${\bf 10}$ representation, but this {\bf 10} only contains
a part of the two light Higgs doublets.
Moreover, we  studied the implications of the down quark mass
corrections, under the assumption that, within a good approximation, 
a super GIM mechanism is in effect. We have shown that, for 
$\omega < 1$ (moderate values of $\tan\beta$), and considering
the simplest operator structure,  
large bottom mass corrections are helpful in 
accomodating the experimentally preferred values for $M_t$, 
yielding also acceptable values for $V_{cb}$. However,
moderate or low values of $\tan\beta$ lead 
to wrong predictions either for the first generation quark masses
or for the CP-odd sector of the theory, a property that is not
changed by the presence of  supersymmetric threshold corrections. 
We have also shown that the operator
structure may be extended to yield proper values for all
fermion masses and mixing angles for low values of $\tan\beta \leq
3$. This extension requires, however, the presence of additional
{\bf 45} states in the theory, one new operator contributing to the first
generation masses and another complex parameter.  

\section{Appendix -- Supersymmetric threshold corrections}

Let us discuss the down quark mass corrections induced by
supersymmetric particle loops in 
more detail. The dominant  corrections to the 
down quark mass matrix are given by chargino--up squark
and gluino--down squark one loop contributions and they read,  
\begin{eqnarray}
\left(\delta m_d \right)_{I L} & = & - \frac{2\alpha_3}{3 \pi}  
M_{\tilde{g}} \sum_{j=2}^{6} \left[
D_{I j} D^{*}_{(L+3) j} (m_{\tilde{d}_j}^2 - m_{\tilde{d}_1}^2)
 I(m_{\tilde{d}_j}^2 , m_{\tilde{d}_1}^2, M_{\tilde{g}}^2)
\right]
\label{eq:bmcorr}
\\
& + &  \sum_{j = 2}^6 \sum_{\alpha = 1}^2 
\left[ \frac{d_L}{16 \pi^2} Z^{+}_{2 \alpha} 
Z^{-}_{2 \alpha} m_{\alpha}
 C^{*}_{K L} U^{*}_{K j}  u_M  U_{(M+3) j} C_{M I}
(m_{\tilde{u}_j}^2 - m_{\tilde{u}_1}^2)
I(m_{\tilde{u}_j}^2,m_{\tilde{u}_1}^2,m_{\alpha}^2) \right].
\nonumber
\end{eqnarray}
The above expression has also been obtained in Ref. \cite{SS}. 
In the above,
$U$ and $D$  are the unitary matrices diagonalizing the six by six 
up and down squark
mass matrices, ($D_{1i}$ and $D_{4i}$ denote, for example,
 the component of the mass
eigenstate $\tilde{d}_i$ in the left and right handed 
down squark, respectively),
$Z^{\pm}_{\alpha \beta}$ are the unitary matrix which diagonalize the two by
two chargino matrix, $m_{\alpha}$ are the chargino mass eigenstates,
$C_{I J}$ are the CKM matrix elements,  $d_I$ and $u_I$ 
are the down
and up quark Yukawa couplings, respectively, and $M_{\tilde{g}}$
is the gluino mass.  The integral $I(a,b,c)$ is given by
\begin{equation}
I(a,b,c) = \frac{ a b \ln(a/b) + b c \ln(b/c) + a c \ln(c/a)}
{(a - b)(b - c)(a - c)}. 
\end{equation}
All indeces denoted by capital letters run from 1 to 3 
and a summation over the indeces $K$ and $M$  is implicit. 
The state $\tilde{d}_1$ 
($\tilde{u}_1$) denotes
any particular eigenstate, which may be chosen, for example, as the heaviest 
one. A dependence
of the above expression on the quark mass matrices is implicit in the 
necessary left right mixing term, which is only generated by terms
proportional to the quark masses.  

Furthermore, if 
the up and down squark matrices are three
by three block diagonal, implying the existence of a super GIM
mechanism in the theory, the following property is  fulfilled
\begin{equation}
D_{K j}  D^{*}_{(M+3) j} = \pm \delta_{KM}
\frac{d_K (A_{d_K} - \mu \tan\beta) v_1}
{m_{\tilde{d}_K}^2 - m_{\tilde{d}_{(K+3)}}^2}, \;\;\;\;\;\;\;\;
U_{K j}  U^{*}_{(M+3) j} = \pm \delta_{KM}
\frac{u_K (A_{u_K} - \mu \cot\beta) v_2}
{m_{\tilde{u}_K}^2 - m_{\tilde{u}_{(K+3)}}^2}, 
\end{equation}
where $A_K$ are the conventionally defined trilinear
soft supersymmetry breaking terms and $\mu$ is the supersymmetric
Higgs mass parameter appearing in the superpotential. The
positive sign in the above expression corresponds to the case
$j = K$, while the negative sign corresponds to the case $j =K + 3$.
Keeping the dominant terms in the large $\tan\beta$ regime,  
the down quark mass corrections take, hence, a very simple form,
\begin{eqnarray}
\left(\delta m_d \right)_{I L} & = & \frac{2\alpha_3}{3 \pi}
\delta_{IL} (d_L v_1) \tan\beta \; \mu  M_{\tilde{g}}  
 I(m_{\tilde{d}_L}^2 , m_{\tilde{d}_{(L+3)}}^2, M_{\tilde{g}}^2)
\nonumber\\
& + &  \sum_{j = 2}^6 \sum_{\alpha = 1}^2 
\left[
\frac{(d_L v_1) \tan\beta}{16 \pi^2} Z^{+}_{2 \alpha} Z^{-}_{2 \alpha}
m_{\alpha}
C^{*}_{M L}   |u_M|^2 A_{u_M} C_{M I}
I(m_{\tilde{u}_M}^2,m_{\tilde{u}_{(M+3)}}^2,m_{\alpha}^2)
\right] .
\label{eq:dmcorr}
\end{eqnarray}
The above expression reproduces the one obtained in Ref. \cite{SS}
under similar assumptions. In the present limit,
the gluino corrections affect only the values of the mass
eigenstates, while the chargino corrections give also 
corrections to the off--diagonal terms. Studying the renormalization
group evolution of the soft supersymmetry breaking mass parameters
one can show that the gluino contributions are generally dominant
and opposite in sign to the chargino contributions \cite{wefour}.
Moreover, due to the hierarchy between the up quark masses, 
only the term proportional to $|u_3|^2$ becomes important in
the chargino contributions. Hence the 
chargino-induced corrections to the down and strange  masses are
very small. Furthermore, as has been shown in Ref. \cite{SS},  the 
most relevant corrections to the CKM matrix elements
are given by $\delta V_{c b}/V_{cb}
\simeq - (\delta m_b/ m_b)^{ch.} \simeq \delta V_{td}/V_{td}$, 
where $(\delta m_b/ m_b)^{ch}$ represents only
the  chargino contributions to the total bottom
mass corrections.  
~\\

\begin{figure}
\centerline{
\psfig{figure=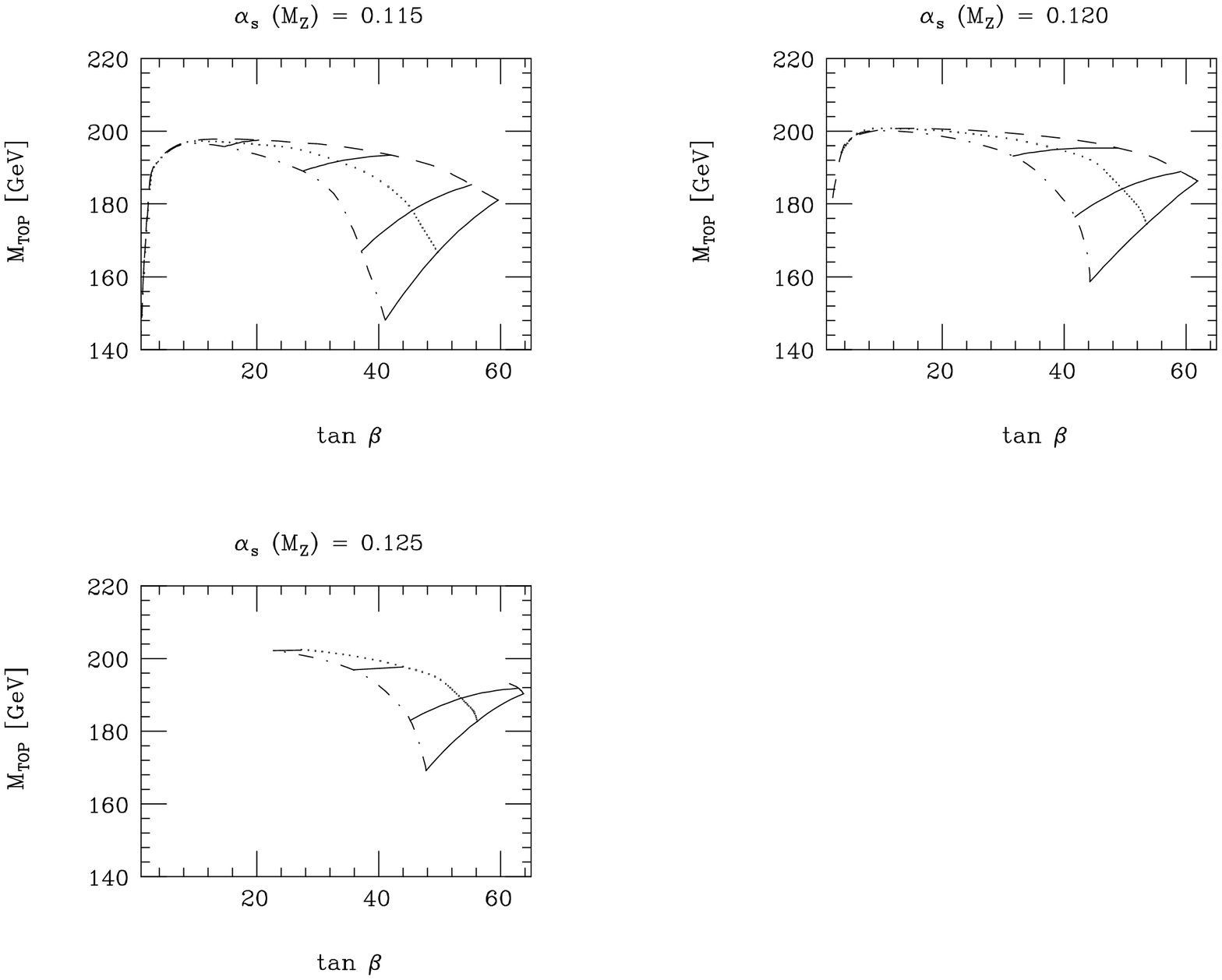,width=20cm,height=15cm,angle=90} }
\caption[0]
{a) The pole top quark mass as a function of $\tan\beta$ for 
a running bottom mass $m_b(m_b) = 4.15$ GeV and three different 
values of the strong gauge coupling, $\alpha_s(M_Z) = 0.115$,
0.120 and 0.125, respectively, for model 6. 
The coefficient $K_c$ 
parametrizing the down quark mass corrections takes values,
$K_c = 0$ (dashed line),
$K_c = -0.003$ (dotted line) and $K_c = -0.006$ (dot-dashed
line). The solid lines  
represent, from right to left, values of $\omega = 1$,
0.6, 0.2 and 0.06, respectively. For large values of 
$\alpha_3(M_Z)$ the curves are cutted at the point at which
the top Yukawa coupling becomes strong at high energy 
scales, $h_t^2(M_{GUT})/4\pi \geq 1$.}
\end{figure}
\setcounter{figure}{0}
\begin{figure}
\centerline{
\psfig{figure=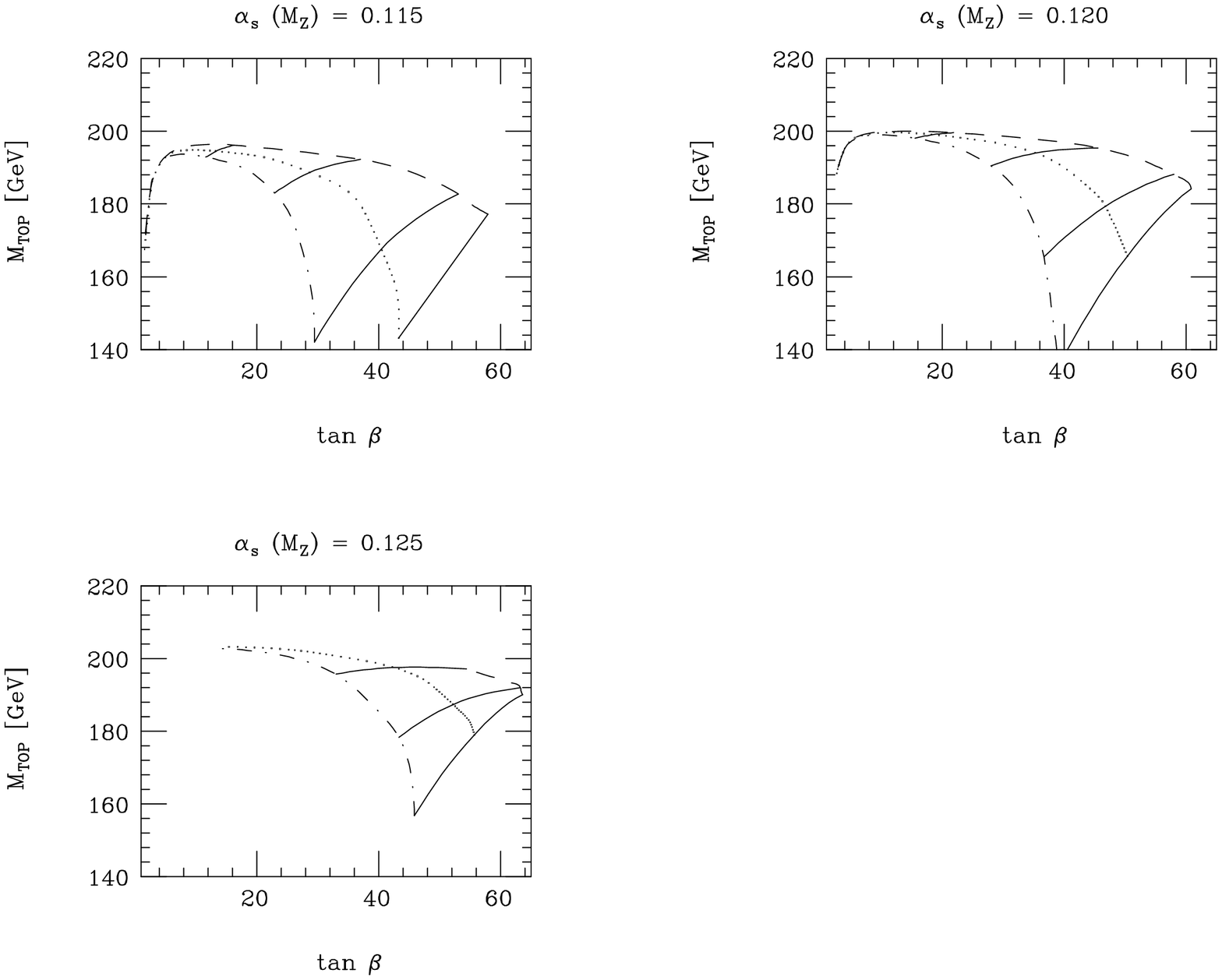,width=20cm,height=15cm,angle=90} }
\caption[0]
{b) The same as Fig. 1.a but for model 9.}
\end{figure}
\begin{figure}
\centerline{
\psfig{figure=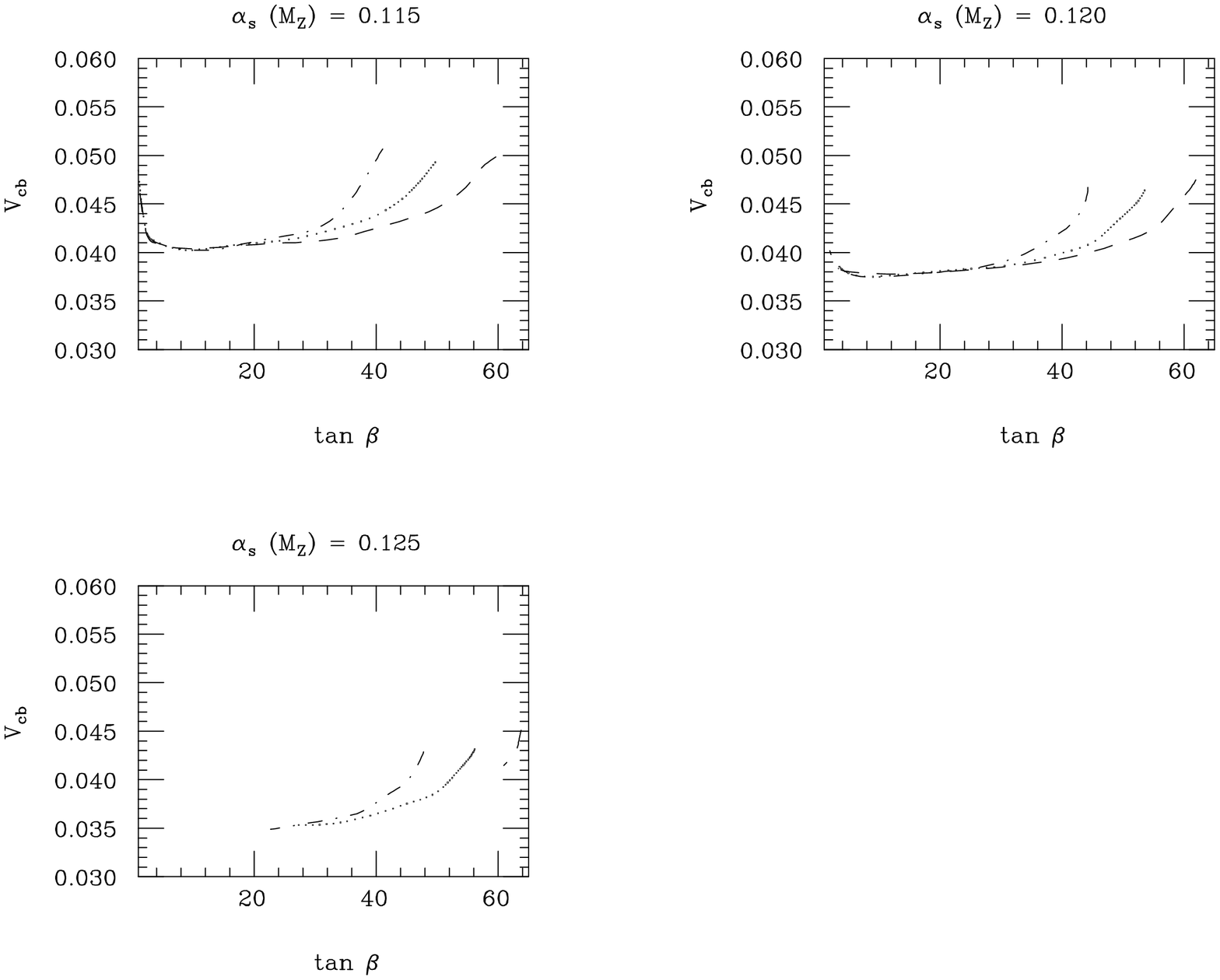,width=20cm,height=15cm,angle=90} }
\caption[0]
{a) The same as in Fig. 1.a but for 
the Cabibbo-Kobayashi-Maskawa matrix element $V_{cb}$ as a function
of $\tan\beta$.}
\end{figure}
\setcounter{figure}{1}
\begin{figure}
\centerline{
\psfig{figure=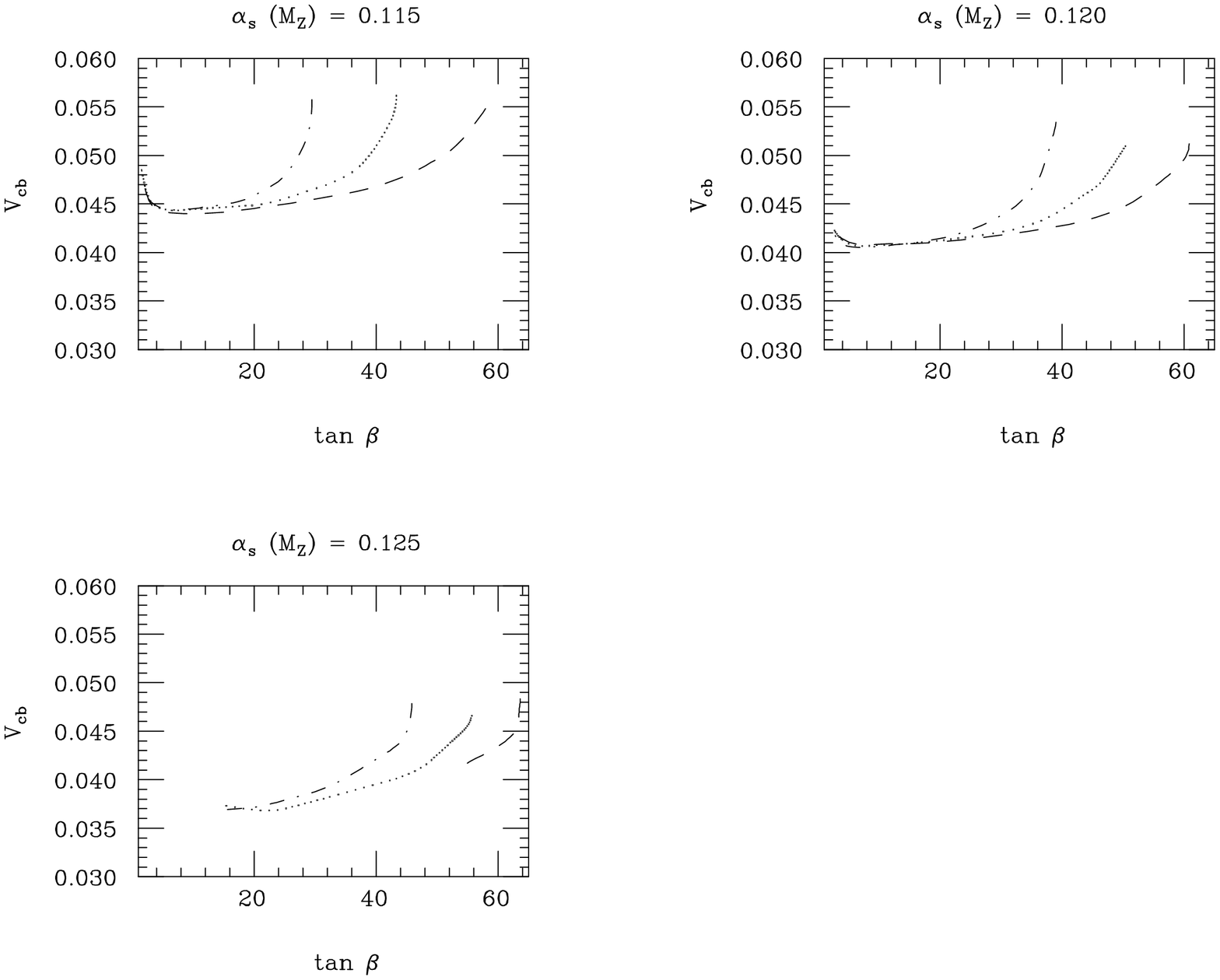,width=20cm,height=15cm,angle=90} }
\caption[0]
{b) The same as in Fig. 2.a but for model 9.}
\end{figure}
\begin{figure}
\centerline{
\psfig{figure=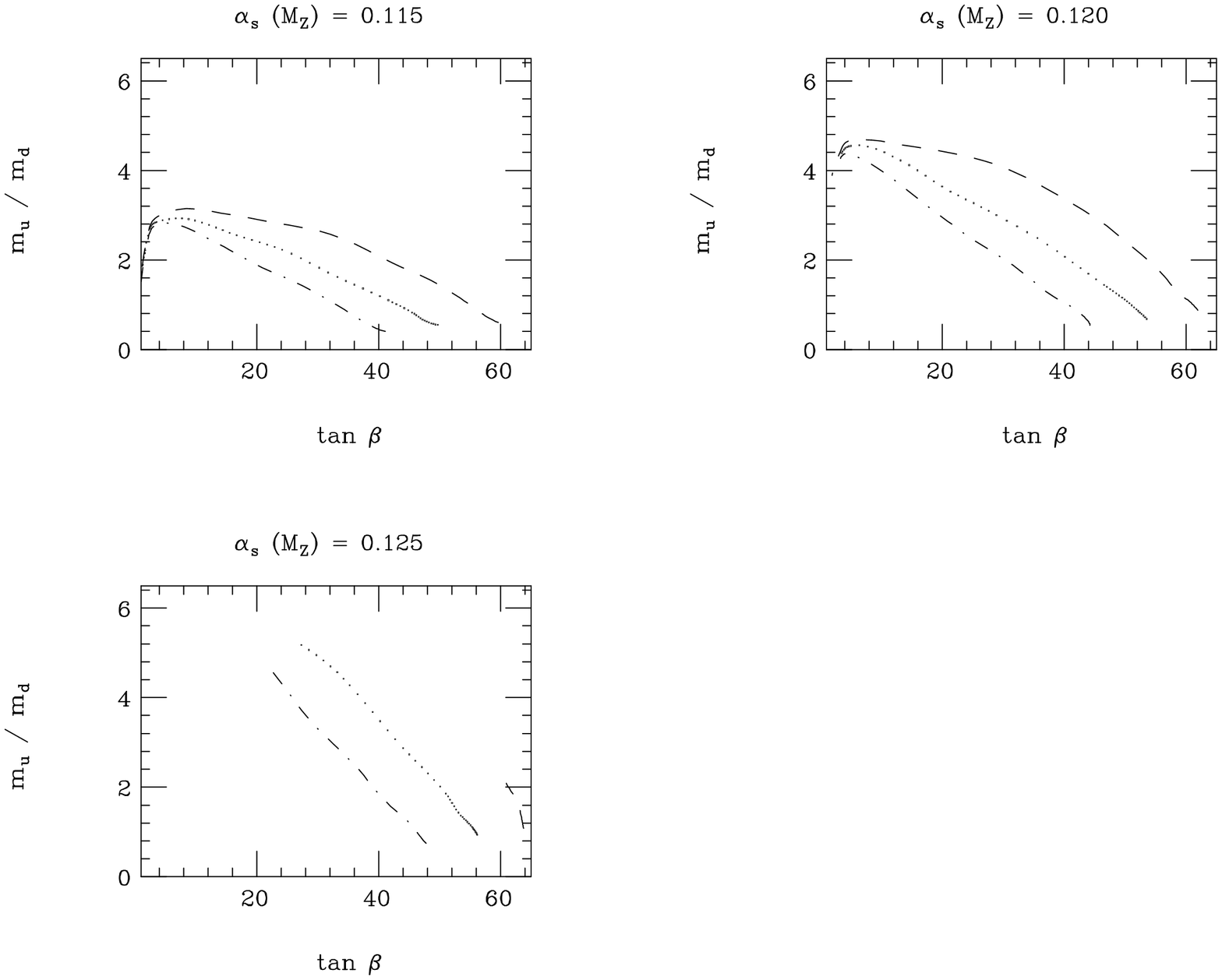,width=20cm,height=15cm,angle=90} }
\caption[0]
{a) The same as in Fig. 1.a but for the ratio of the
first generation masses $m_u/m_d$ as a function of $\tan\beta$.}
\end{figure}
\setcounter{figure}{2}
\begin{figure}
\centerline{
\psfig{figure=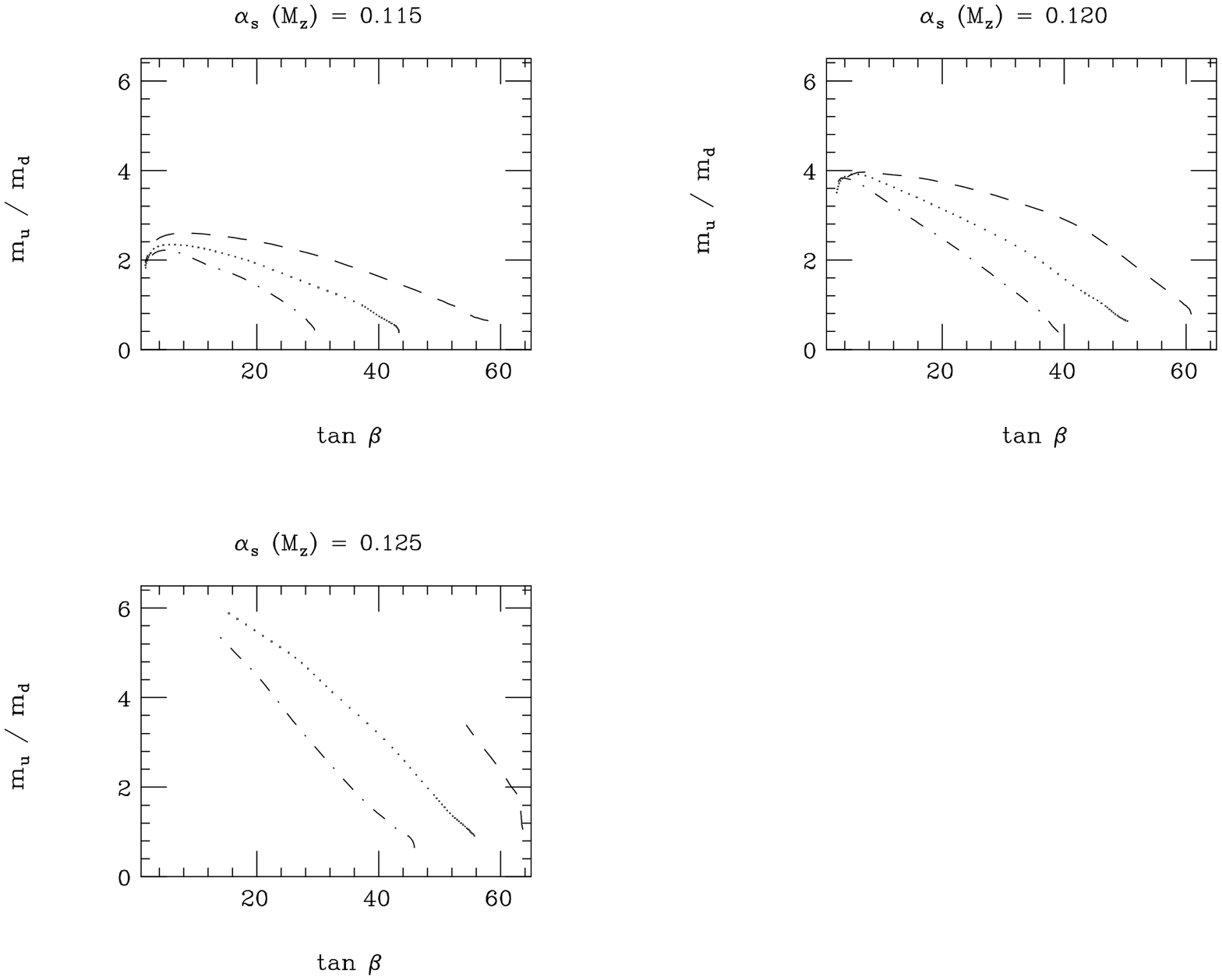,width=20cm,height=15cm,angle=90} }
\caption[0]
{b) The same as in Fig. 3.a but for model 9.}
\end{figure}
\end{document}